\documentclass[12pt]{amsart}
\usepackage{latexsym,fancyhdr,amssymb,color,amsmath,amsthm,graphicx,listings,comment}
\usepackage[section]{placeins}
\let\paragraph\subsection

\title{Remarks about the Arithmetic of Graphs}
\author{Oliver Knill}
\date{June 17, 2021, updated July 18, 2021}
\address{Department of Mathematics \\ Harvard University \\ Cambridge, MA, 02138 }
\subjclass{05C25,  
           46J10,  
           13Axx,  
           68R10}  
\keywords{Graph theory, Arithmetic, Calculus on graphs}

\begin{document}
\maketitle

\begin{abstract}
The arithmetic of $\mathbb{N} \subset \mathbb{Z} \subset \mathbb{Q} \subset \mathbb{R}$ 
can be extended to a graph arithmetic $\mathcal{N} \subset \mathcal{Z} \subset \mathcal{Q} \subset \mathcal{R}$, where 
$\mathcal{N}$ is the semi-ring of finite simple graphs and where $\mathcal{Z},\mathcal{Q}$ are integral domains 
culminating in a Banach algebra $\mathcal{R}$. An extension of $\mathbb{Q}$ with a single network 
completes to the Wiener algebra $A(\mathbb{T})$. We illustrate the compatibility with topology and spectral theory.
Multiplicative linear functionals like Euler characteristic,  the Poincar\'e polynomial, 
zeta functions can be extended naturally. These functionals can also help with number theoretical questions.
The story of primes is a bit different as the new integers $\mathcal{Z}$ are not
a unique factorization domain, because there are many additive primes and because a simple sieving argument shows
that most graphs in $\mathcal{Z}$ are multiplicative primes unlike in $\mathbb{Z}$, where most are not. 
\end{abstract}

\section{Graph arithmetic}

\paragraph{\large{Overview and the beginnings of graph arithmetic.}}
The foundations of {\bf graph arithmetic} have been led in the 1950'ies by Claude Shannon \cite{Shannon1956},
Gert Sabidussi \cite{Sabidussi} and Alexander Zykov \cite{Zykov}. 
The subject has flourished in graph theory \cite{ImrichKlavzar,HammackImrichKlavzar}. 
We got interested in the subject in \cite{AdventuresAlgebra,StrongRing, ArithmeticGraphs,numbersandgraphs}. 
Addition and multiplication of finite simple graphs $\mathcal{N}$ leads to expressions like $(X^2+2X+3)(Y^3+1)$. Examples of rational 
networks like $X+4X^2-3/X$ are obtained by localizing the polynomial ring $\mathcal{Z}$ to a Laurent ring $\mathcal{Q}$.
For expressions like $\cos(X)=1-X^2/2!+X^4/4!-\dots, \log(1+aX)=1-aX/1+a^2X^2/2-\dots$ or rational expressions $1/(1-aX)=1+aX+a2X^2+\dots$ 
an analytic functional calculus is needed for taking limits of polynomial values. There is a completion $\mathcal{R}$ of a 
localization $\mathcal{Q}$ of the ring of integers $\mathcal{Z}$ which itself was a group completion
of $\mathcal{N}$. Completing $\mathcal{Q}$ to a Banach algebra $\mathcal{R}$ is the best we can hope for because
of theorems of Frobenius, Hurwitz, Gelfand and Mazur that classify division algebras and imply that
a norm satisfying $|A*B| = |A| \cdot |B|$ can not exist on the ring $\mathcal{Z}$. 
For any multiplicative non-negative 
functional $c: \mathcal{N} \to \mathbb{R}$ for which $c(A)=0$ implies $A=0$ one gets a {\bf norm ring structure} 
on $\mathcal{Z}$ by defining $|X|$ as the minimum of $|A-B|=c(A)+c(B)$ for $X=A-B$ and extend it by $|X A^{-1}|=|X|/|A|$ for 
$A \in \mathcal{N}$ to $\mathcal{Q}$ satisfying still the inequality $|X*Y| \leq |X| |Y|$ then complete 
this {\bf normed ring} to the Banach algebra $\mathcal{R}$. The simplest and best choice is $c(A)=f_0(A)$ the number of vertices
in $A$ because we want that on the usual rational numbers $\mathbb{Q}$,
the norm $|A|$ is the same than the usual norm. The choice of {\bf localization} $\mathcal{Z} \to \mathcal{Q}$ is done so
that the normed ring property extends to $\mathcal{Q}$ and so will allow for a completion $\mathcal{R}$ containing the usual
real numbers $\mathbb{R}$. 

\paragraph{\large{Notation and nomenclature.}}
The set $\mathcal{N}$ of graphs is always assumed to consist of
finite simple graphs. The word ``graph"  is also be used for the larger set of integers $\mathcal{Z}$, signed graphs
of the form $A-B$, where $A,B$ are both graphs. We can assume that $A,B$ to have no common non-zero 
additive component. The elements $(A+C) - (B+C)$ and $A-B$ represent the same integer.
While in $\mathbb{Z}$, we can always write $A$ or $-A$ for any integer, this is not possible in $\mathcal{Z}$.
As we always deal with graphs, the algebra generated by $0$-dimensional graphs is identified with the traditional field
of real numbers $\mathbb{R}$. It is embedded in $\mathcal{R}$ so that $\mathcal{R}$ will be a real Banach algebra.
We could extend that field to $\mathbb{C}=\mathbb{R}[i]$ to get completeness among zero
dimensional graphs. The space $\mathcal{Q}$ will be a localization using the multiplicative monoid generated by
non-zero connected components in $\mathcal{N}$ and not the ring localized at all non-zero graphs $\mathcal{Z}$.
The later would produce a field, the field of fractions, but such an extension would not allow
the norm $\mathcal{Z}$ to be continued. Like the polynomial ring $F[X]$ which when localized at $S=\{ X^n, n \geq 0 \}$
becomes $F[X,X^{-1}]$, the smallest ring in which $X$ is a unit and where the field of fraction $F(X)$ of $F[X]$ is much larger.
With  ``integer" we mean an element in $\mathcal{Z}$ while elements in $\mathbb{Z}$ are called 
``rational integers".

\paragraph{\large{What is new here?}}
This is our fourth approach to graph arithmetic.
Unlike in \cite{AdventuresAlgebra,StrongRing, ArithmeticGraphs,numbersandgraphs} we now favor to work with the Shannon ring and not the 
isomorphic Sabidussi ring. The only really new thing in this note is to have a better sense about completion. Especially 
the Wiener picture for a one-network extension helps because this goes much beyond an analytic functional calculus.
In an analytic functional calculus, we have an integral domain property $f(G) g(G)=0$ implying $f(G)=0$ or $g(G)=0$. Already in the 
Wiener algebra this is no more true. We can find $f(G) g(G)=0$ without both $f(G)$ and $g(G)$ being zero. 
Our perception of graph arithmetic has considerably shifted in that we think more and more of graphs as numbers. 
We have pointed out already earlier the possibility to extend multiplicative quantities to the Banach algebra. 
Multiplicative linear functions like the Euler characteristic $X$, the Poincar\'e polynomial $p$ or    
various zeta functions can be extended naturally to analytic elements in $\mathcal{R}$. 
The Poincar\'e poynomial $p_G(t)$ of a graph for example is 
completely compatible with arithmetic as K\"unneth shows. The K\"unneth formula is explicit
in that we can construct the harmonic forms of a product in terms of the harmonic forms of the factors.
Also cohomology extends using functional calculus:
the k'th coefficient of $\exp(p_G(t))$ for example can be seen as the $k$'th Betti number of $\exp(G) \in \mathcal{R}$ and 
$2^{\chi(G)})$ is naturally the Euler characteristic of $2^G \in \mathcal{R}$ defined by making a Taylor expansion of the 
function $f(x)=2^x$. The setup is motivated also from the concept of {\bf Shannon capacity} $\limsup_{n \to \infty} 
\alpha(G^n)^{1/n}$ of $G$ which involves the arithmetic of graphs. We can now see this capacity also in the context of calculus:
what is the radius of convergence of a function $f$ have to be to make sense of the independence number 
$\alpha(f(G))$. 

\paragraph{\large{The strong ring of Shannon.}}
The graph multiplication of Shannon \cite{Shannon1956} is today mostly called the {\bf strong multiplication}. Formally, it is \\
$(V,E) \star (W,F)=(V \times W,\{ ((a_1,a_2),(b_1,b_2)), (a_1,b_1) \in E \cup \{ (v,v), v \in V\}, 
                                                         (a_2,b_2) \in F \cup \{ (v,v), v \in W\} \}$. 
This associative multiplication with the $K_1$ as the $1$ elements leads after a Grothendieck completion of the additive 
monoid $\mathcal{N}$ of finite simple graphs with disjoint union as addition to the {\bf commutative unital ring} 
$\mathcal{Z}$ in which the {\bf disjoint union} is the addition and the {\bf strong graph product} is the multiplication.
When graphs are restricted to $0$-dimensional graphs (graphs without edges), we have the usual integer
arithmetic $\mathbb{Z}$, where the disjoint union is the addition and the Cartesian product the multiplication. 
Unlike for $\mathbb{Z}$, where every number can be written either as $X=A$ or $X=-A$, in 
$\mathcal{Z}$, we must represent a number in the form $X=A-B$ in general. The reason is the existence of 
{\bf additive primes} in $\mathcal{N}$ which are the connected graphs. This is richer than in $\mathbb{N}$, 
where we have only $1$ as an additive prime. The ring $\mathcal{Z}$ will turn out to be an integral domain and
already leads to interesting problems like characterizing multiplicative primes or determining {\bf how costly it is}
to factor a general connected graph into prime graphs. (For connected graphs the factorization is unique.)

\paragraph{\large{Extending the vision of Shannon.} }
The following verbatim quote from the pioneering work \cite{Shannon1956} motivates the topic:
{\it "The sum of channels corresponds physically to a situation where either of two channels may be
used (but not both), a new choice being made for each transmitted letter. The product channel
corresponds to a situation where both channels are used each unit of time.  It is interesting
to note that multiplication and addition of channels are both associative and commutative,
and that the product distributes over a sum. Thus one can develop a kind of algebra for channels in 
which is possible to write, for example, a polynomial, where the $\sum_n a_n K^n$, where
the $a_n$ are non-negative integers and $K$ is a channel. We shall not, however, investigate
here the algebraic properties of this system."}
We might add also that in initially like in \cite{AdventuresAlgebra,StrongRing, ArithmeticGraphs} we have first
not been aware yet of Shannon's paper in graph arithmetic. 
The statement ``thus one can develop a kind of algebra for channels" of Shannon can be extended to 
``Thus one can develop a calculus for channels". Having a completed algebra $\mathcal{R}$ allows for example to  
consider {\bf waves of graphs} by looking at $\exp(i G t) = \cos(Gt) + i\sin(Gt)$, where $G$ is a graph.

\paragraph{\large{The large ring of Sabidussi.}}
The Shannon ring is an {\bf integral domain} isomorphic to the {\bf Zykov-Sabidussi ring}
\cite{Zykov,Sabidussi} in which the {\bf Zykov join} is the addition and the {\bf large multiplication} is the product. 
The Zykov join (or simply {\bf join}) $A \oplus B$ of two graphs is the disjoint union $A+B$ modified by additionally joining
every vertex in $A$ with every vertex in $B$. The Sabidussi multiplication has like the strong multiplication the Cartesian
product as vertex set. Two grid points are now connected, if one of the projections is a vertex or edge. 
The isomorphism is given a {\bf natural graph complement endomorphism} $G \to \overline{G}$ in $\mathcal{N}$ 
which maps a graph to its {\bf graph complement}. This endomorphism is an involution that 
maps $0$-dimensional graphs $\mathbb{N}$ to complete graphs: $\overline{P_n}=K_n$ and $\overline{K_n} = P_n$. 
It maps cyclic graphs $C_n$ to graphs which are homotopic to spheres or wedge sums of 
spheres and linear graphs $L_n$ to graphs with interesting curvature universality \cite{GraphComplements}.
Because of this natural symmetry in graph theory, the two pictures are isomorphic. We only need to consider one of them.
The Shannon picture with the strong multiplication is closer in nature to the Cartesian product for topological spaces.
It is also more intuitive as it is the hybrid of a small product and a tensor product. Still, the dual picture can 
be interesting too, considering that spheres are preserved by joins. 

\paragraph{\large{Algebraic ring of Stanley-Reisner.}}
The Shannon ring is not only dual to the Sabidussi ring. It is also related to the {\bf Stanley-Reisner ring}. 
If $A$ is a graph with vertex set $V=\{v_1, \dots, v_n\}$, the {\bf Stanley-Reisner polynomial}
$s_A(v) = s_A(v_1,\dots,v_n)$ is a sum over all {\bf monomial expressions} $v_{i_1} \cdots v_{i_k}$, where we sum over complete
subgraphs $(v_{i_1},\dots,v_{i_k})$ of $A$. The polynomial product $s_{A B}(v,w)=s_A(v) s_B(w)$ defines a new graph
in which the {\bf monomial expressions} of the polynomial are the vertices and two monomials are connected if 
one divides the other. This new graph is called the {\bf Barycentric refinement} of the product of the 
{\bf Whitney complexes} of $A$ and $B$. (The Whitney complex is the finite abstract simplicial complex containing as
sets the vertex sets of complete subgraphs. The Cartesian product as sets of sets is not yet an abstract simplicial 
complex any more because it is not closed under the operation of taking finite subsets).
If we take the Stanley-Reiser polynomial $s_A$ of $A$ and look at the {\bf connection graph} $A'$ which again has
the monomials as vertices but where now two monomials are connected if they have a {\bf common non-zero divisor},
then $A' * B' = (A \times B)'$ is the strong product of the connection graphs. 

\paragraph{\large{A tensor representation of the ring.}}
While the {\bf incidence calculus} pioneered by mathematicians like Poincar\'e or Betti on discrete spaces 
and the calculus on manifolds look similar (both coming from an 
exterior derivative $d$), the connection picture is only available in the discrete. 
The corresponding {\bf connection Laplacian} $L$ are invertible and the {\bf Green functions} the matrix entries of
$g=L^{-1}$ are bounded. We can describe the inverse $g$ of the connection Laplacians $L$ with explicit finite expressions. The matrices are
invertible so that all Green functions are bounded. The mechanisms are also close what one sees when looking at mathematical
structures in {\bf quantum mechanics}: for any {\bf energized connection Laplacians} $L_A$
we have $L_{A+B} = L_A \oplus L_B$ and $L_{A*B} = L_A \otimes L_B$ which on the level of spectra means
$\sigma(L_{A+B}) = \sigma(L_A) \cup \sigma(L_B)$ the {\bf union of the spectra} and 
$\sigma(L_{A*B}) = \sigma(L_A) \sigma(L_B)$ the {\bf product of the spectra} \cite{KnillEnergy2020,GreenFunctionsEnergized}. 
The fact that the strong graph product is related to a tensor product of
matrices is relevant if one looks at {\bf representation theoretical aspects}. As we will see below, it is very useful
because we can construct {\bf explicit harmonic forms} when taking products and illustrate the cohomology ring
as the {\bf cup product} is easier to implement. This is how Hassler Whitney thought about
the cup product \cite{WhitneyCollected}. We come to this next.

\paragraph{\large{Shannon and Stanley-Reisner products as cousins in a Cartesian product framework.}}
Let us dwell a bit more on seeing both the Shannon and Stanley-Reisner product as a natural ``Cartesian product".
We can actually see them just different manifestations and equivalent if we identity homotopic graphs.
If $A_1,B_1$ are the Barycentric refinements of graphs $A,B$, then
one can interpret the Stanley-Reisner product $(A \times B)_1$ as a {\bf Barycentric refinement} of the virtual
$A \times B$, even so $A \times B$ is not defined as a graph (one classically uses CW complexes to handle such things). 
It is important that while $A \times B$ itself just
has only an algebraic meaning at first, the graphs $(A \times B)_1$ and $(A \times B)'$ are both well 
defined and both share the properties which we want to have for a Cartesian product. A nice thing about $(A \times B)'$ is
that it preserves discrete manifolds.  If $A',B'$ are the connection graphs, then the Shannon product 
satisfies $A' * B' = (A \times B)'$. The Stanley-Reisner product is just an ``incidence version" of the Cartesian
product and the Shannon product gives a ``connection version of the Cartesian product".
While connection graphs $A'$ have in general higher dimension than $A$, this is not a draw-back:
the connection graph $A'$ and the Barycentric refinement graph $A_1$ are homotopic of $A$ is already refined. 
\cite{ComplexesGraphsProductsShannonCapacity}. 

\paragraph{\large{On cohomology and cup product.}}
The {\bf K\"unneth formula} (see \cite{KnillKuenneth} in the Stanley-Reisner picture) is much easier to see in 
the Shannon product picture because if $f$ is a $k$-form on $A$ and $g$ is a $l$-form on $B$ then $d^*(f \otimes g)$ 
is a $(k+l)$- form on $A*B$. The Hodge Laplacian $H_{A'*B'}$ acts on the differential forms of the connection graph $A'*B'$. 
The harmonic forms of $A' * B'$ are all composed of harmonic forms in $A'$ and harmonic forms in $B'$. 
(we need to scale down with $d^*$ because if $f$ is a function of $p+1$ variables and $g$ is a function of $q+1$ variables
then the tensor product $f \otimes g$ is a function of $p+q+2$ variables which is one too much. By applying a divergence $d^*$
we get a function of $p+q+1$ variables and so a $p+q$-form for which it is easy to see that if both $f,g$ are harmonic then 
$d^*(f \otimes g)$ is harmonic on
$A' \star B'$.  As the connection graph $A'$ has the same cohomology than $A_1$ if $A$ was Barycentric refined, the
K\"unneth formulas does not need a {\bf chain homotopy argument} as translated in \cite{KnillKuenneth} to the discrete.
(An alternative is to use discrete CW complexes and use cellular cohomology but CW complexes are harder to implement in 
a computer.) The {\bf homotopy} argument  going from the incidence to the connection graph makes
harmonic forms  explicit and harmonic forms are cohomolgy classes, this directly implements the {\bf cup product}.
It is nothing else than the tensor product of forms in the connection graphs on which a $d^*$ was applied.
That no averaging is needed to define the cup product in the discrete has first been realized by Whitney, 
who was one of the first who considered the cup product in cohomology. 
This is really useful as we can implement the {\bf cohomology ring} explicitly in a computer as a 
tensor algebra. This is done by looking at the equivalent cohomology on $A'*B'=(A \times B)'$ given by Shannon and not
as $(A \times B)_1$ given by the Stanley-Reisner product. The connection picture is more elegant. 

\paragraph{\large{Poincar\'e polynomials.}}
A reformulation of the {\bf K\"unneth formula} is that the {\bf Poincar\'e polynomial map} $A \to p_A(t)$ is a ring 
homomorphisms from $\mathcal{N}$ to $\mathcal{N}[t]$. 
We need to define the cohomology also for $A \in \mathcal{Z}$ 
(the group completion with negative elements too) and not only for $A \in \mathcal{N}$ which are graphs. 
For cohomology, the {\bf Hodge Laplacian}
$H_A = (d+d^*)^2 =D^2$ defined by the exterior derivative $d=d_A$ is the relevant Laplacian. Its {\bf Betti numbers}
are the dimension of the kernels on the blocks of $H_A$. The {\bf exterior derivative} $d_{A \times B}$ 
of the product is also defined for the Cartesian product $(A \times B)_1$. We have just seen that
the cohomology for $(A \times B)'$ can be identified with the cohomology of the Shannon product 
$A'*B'$ of the connection graphs. Both the {\bf f-vector} $f_k(A)$ counting the number of complete
sub-graphs of dimension $k$ as well as the {\bf Betti numbers} $b_k(A)$ can be naturally extended from 
$\mathcal{N}$ to $\mathcal{Z}$ by defining $f_k(-A)=-f_k(A)$ and $b_k(-A)=-b_k(A)$. 
When done so, both the {\bf Poincar\'e polynomial} $p_A(t) = \sum_k b_k(A) t^k$ with {\bf signed Betti numbers} 
$b_k(A)$. Euler polynomial  $e_A(t) = \sum_{k=0} f_k(A) t^k$ with {\bf signed $f$-vector} 
is an additive {\bf group homomorphisms} from the Shannon ring $\mathcal{Z}$ to the polynomial ring $\mathbb{Z}[t]$. 

\paragraph{\large{Entangled states and information.}}
Having at hand a related {\bf tensor algebra} of matrices reminds of similar constructions in {\bf particle physics}, where one has a concept 
of {\bf second quantization} using {\bf Fock calculus}. The tensor product of states encoding {\bf entangled states}
is familiar not only in physics but also in {\bf information theory}. The {\bf Lovasz umbrella} \cite{Lovasz1979} 
for example can be interpreted as the process of attaching a {\bf quantum state} $U(v)$ (unit vector in some vector space) 
to each vertex $v$ of the graph. 
Quantum mechanical states assigned to non-adjacent vertices must be perpendicular
(uncorrelated) to ensure causality. When taking the Shannon product of graphs, the Lovasz umbrellas {\bf tensor multiply}.
This remains true also in the more general frame work of density matrices. The {\bf Lovasz number} 
$\theta(G) = \inf_{U,c} \max_{v} (c \cdot U(x))^{-2}$ (where the {\bf umbrella stick} $c$ is an other unit vector)
is then multiplicative $\theta(G \times H) = \theta(G) \theta(H)$
allowing it to be used as an upper bound for the Shannon capacity $\Theta(G)=\lim_{n \to \infty} \alpha(G^n)^{1/n}$,
where $\alpha(A)$ is the {\bf independence number} of the graph $A$ and $G^n =G*G* \cdots *G$ is the $n$'th power in the Shannon ring.
The story is well told in \cite{Matousek}. See also \cite{ComplexesGraphsProductsShannonCapacity}.

\paragraph{\large{To get to a normed ring}}
The semi-ring $\mathcal{N}$ has a norm $c(X)$ given by the {\bf number of vertices in $X$}.
It is a linear and multiplicative property $c(X+Y)=c(X) c(Y), c(X*Y) = c(X) c(Y)$ and
defines $|X|=c(X)$ for $X \in \mathcal{N}$. On the ring $\mathcal{Z}$ we can define now 
$|X| = {\rm min}_{X=A-B} c(A) + c(B)$. This is $|X|=c(A)+c(B)$ if $A$ collects all positive connected components in $X$
and $B$ contains all negative connected components of $X$.  The norm still satisfies
have $|X*Y| \leq |X| |Y|$ on $\mathcal{Q}$. We get so a {\bf normed ring} \cite{numbersandgraphs}. The choice with $c(X)$ is only one of
many possible choices. An other one would be the total number of simplices in $G$.
Now, one can complete $\mathcal{Q}$ to a Banach algebra $\mathcal{R}$. Unlike in $\mathcal{Q}$, we can do more in a
Banach algebra $\mathcal{R}$. For example, we can define $\log(1+a X)$ for small enough $a$ or $\exp(X)$ using functional
calculus. The Banach algebra $\mathcal{R}$ is not a $C^*$-algebra as the property $|A*A^*| = |A^*| |A|$ can not be achieved.
The choice of norm is far from unique.
One could take the {\bf Kalai number} $c(X) = \sum_k f_k(X)=f_0(X_1)$ counting the number of simplices in $X$.
(We just want the quantity to give the usual norm when restricted to zero dimensional graphs). 

\paragraph{\large{Weighted Wiener algebras.}}
A recent thing here is that we can identify the {\bf single network algebra} $\mathcal{R}[G]$ with a {\bf Wiener algebra}
which is isomorphic to the classical Wiener algebra $A(\mathbb{T}) \sim l^1(\mathbb{Z})$ if the norm is chosen so that $|G|=1$.
The product algebra $\prod_{P} \mathcal{R}[P]$ is only part of the full algebra. These are the {\bf pure states}. There also
expressions like $X^2+B^2$ which can not be written as a product. These are {\bf entangled states}. 
The choice of $c(G)=|G|$ on integers tells how the norm on the $1$-particle Wiener algebras is weighted. In the ring $\mathcal{R}[G]$
generated by one network $G$, the norm is $|G|=\sum_{n \in \mathbb{Z}} |a_n| c(G)^n$. Whatever choice has been done for $c(G)$, the
full ring $\mathcal{R}$ is huge. It contains the Laurent polynomial rings of any number of variables and product rings and these
product rings generate the full algebra. 
Each {\bf connected prime graph} $G \in \mathcal{P}$ already defines a fresh new variable. The ring generated by one variable 
$G$ then contains then as rational numbers elements of the form $\sum_{k=-l}^m a_k G^k$ which for $l \geq 0$ are 
in $\mathcal{Z}$ and otherwise are in $\mathcal{Q}$. Unlike in abstract algebra where $\sum_{k=-l}^m a_k x^k$ 
is just a Laurent polynomial, we have in our case geometric objects $G^n$ as well as more formal versions $-G^n$ or
$G^{-n}$. Any geometry $G$ in $\mathcal{Q}$ comes now with a {\bf cohomology ring} $H^k(G)$, 
a {\bf tensor algebra of connection matrices} $L_G$ and
the related {\bf zeta function} $\zeta_A(s)$. 

\paragraph{\large{Criteria for a network to be prime.}}
Here is an application on how the arithmetic on $\mathcal{Z}$ can fit with geometric concepts.
We recall first that both the {\bf Euler polynomials} and {\bf Poincar\'e polynomials} are
{\bf ring homomorphisms} $\mathcal{Z} \to \mathbb{Z}[t]$. This 
can be even be extended to $\mathcal{Q} \to \mathbb{Q}(t)$.  For the Euler polynomial, one can see that just by
noting that a $k$-simplex in $A$ and a $l$-simplex in $B$ defines a $(k+l)$-simplex in $A*B$. 
For the Poincar\'e polynomial, the fact that it is a multiplicative linear functional is essentialy rephrasing the 
K\"unneth formula. In particular, the {\bf Euler characteristic} $\chi$ 
extends to $\mathcal{Z}$ and still satisfies {\bf Euler-Poincar\'e relation} $p_A(-1) = e_A(-1)$ for Euler characteristic.
Here is an application to the number theory in $\mathcal{N}$: graphs $A$ with
{\bf irreducible} Euler polynomial $e_A(t)$ are definitely prime. The proof goes by noticing first that
$p_A(t)=0$ implies $A=0$ and $e_A(t)=0$ implies $A=0$ and that 
$e_A(t)=1$ implies $A=1$ if $A \in \mathcal{N}$. Now if $G=A*B$, we can apply the Euler polynomial $e$ on both 
sides and get not zero: $e_{A *B}(t) = e_A(t)* e_B(t)$. The irreducibility assumption now assures that $e_A(t)=1$ or
$e_B(t)=1$. Therefore, $A=1$ or $B=1$. For the Poincar\'e polynomial, we still could have non-prime networks $A*B$,
where $A$ has a trivial cohomology and so $p_A(t)=1$. Cohomology can be useful when considering a {\bf weak prime condition} 
like that $G=A*B$ implies that $A$ has trivial cohomology or $B$ has trivial cohomology.

\paragraph{\large{Wu characteristic and zeta functions}}
For a complete subgraph $x$ in a graph $G$, define $\omega(x) = (-1)^{\rm dim}(x)$, where the dimension ${\rm dim}(x)$ is 
one less than the number of vertices in the simplex $x$. The multiplicative nature $\chi(A*B) = \chi(A) \chi(B)$ 
of the {\bf Euler characteristic} $\chi(G) = \sum_{x} \omega(x)$ is already evident 
when looking at the simplices. Also the more general {\bf Wu characteristic} 
$\omega_k(G)= \sum_{x_1 \sim \dots \sim x_k} \omega(x_1) \omega(x_2) \cdots \omega(x_k)$ summing over all intersecting
simplices $x,y$, generalizing the Euler characteristic $\chi(A)=\omega_1(A)=\sum_x \omega(x)$ is multiplicative. 
An other exciting source of numbers are {\bf zeta functions} $\zeta_A(s) = \sum_{j} \lambda_j^{-s}$ 
defined by the eigenvalues $\lambda_j$ of $L_A^2$. This is defined first for $A \in \mathcal{N}$, then extended to
$A \in \mathcal{Z}$ and then to $\mathcal{Q}$. It defines a ring homomorphism from $\mathcal{Q}$
to the space of entire functions on $\mathbb{C}$. The reason is that if $\lambda_j$ are the eigenvalues of $L_A$ for a graph $A$ and 
$\mu_k$ are the eigenvalues of $L_B$ for a graph $B$, then $\lambda_j \mu_k$ are the eigenvalues of $L_{A*B}$. 
This implies $\zeta_{A*B}(s) = \zeta_A(s) \zeta_B(s)$. Clearly also $\zeta(A+B)(s) = \zeta_A(s) + \zeta_B(s)$ for all complex
numbers $s$. We have defined $\zeta_A(s) = \sum_j (\lambda_j^2)^{-s}$ using a square because $\lambda_j^2$ are then all positive.
These spectral zeta functions become {\bf entire functions} in $s$. (If the operator $L_A$ is compared with the Dirac operator
$D=\frac{d}{dx}$ on $\mathbb{T}$ which has eigenvalues $\lambda_j=n$, then the analogue spectral zeta function is $\sum_n n^{-2s}$.
Unlike for manifolds we do not have to discard the zero eigenvalue in the connection Laplacian picture).
There is substantial choice in building the zeta function because we can modify the connection 
Laplacians by energizing it by attaching numerical values 
$h(x)$ to $k$-dimensional complete subgraphs $x$ of $G$ defining then the Laplacians $L_A(x,y) = \chi(S^-(x) \cap S^+(x))$, 
where $\chi(A)$ for a subgraph is $\sum_{x \subset A} h(x)$, summing over all simplices in $A$. 

\paragraph{\large{The set of signed graphs are the new integers.}}  
Let us just summarize in other words how the integral domain $\mathcal{Z}$ is defined.
The set $\mathcal{N}$ of {\bf finite simple graphs} is a monoid with 
respect to {\bf addition} given by the {\bf disjoint union} $A+B$ of graphs. 
The {\bf additive primes} in $\mathcal{N}$ are the {\bf connected graphs}. 
The sub-monoid of $0$-dimensional graphs $\mathbb{N}$ represent the {\bf natural numbers}
in which $1=K_1$ is the only {\bf additive prime}. Connected graph are the additive primes in $\mathcal{N}$.
The {\bf strong product} $A * B$ of two graphs produces a {\bf semi-ring} $\mathcal{N}=\mathcal{Z}^+$
which contains the $\mathbb{N}$ as a sub semi-ring.
The vertices $V(A * B)$ is the Cartesian product of $V(A)$ and $V(B)$. The edge set $E(A*B)$ contains all
$((a_1,b_1), (a_2,b_2))$ for which both $(a_1,a_2) \in E(A) \cup \{ (a,a), a \in V(A) \}$  
as well as $(b_1,b_2) \in E(A) \cup \{ (a,a), a \in V(A) \}$. 
The additive monoid $\mathcal{N}$ group-completes to the {\bf integers} $\mathcal{Z}$ with 
$\mathcal{N}$ being an additive sub-monoid of $\mathcal{Z}$. Now $\mathcal{Z}$ becomes a 
{\bf commutative associative ring} with {\bf $0$-element} $0$ and {\bf $1$-element} $K_1$.

\paragraph{\large{Additive and multiplicative primes.}}
All elements in $\mathcal{Z}$ can be written with {\bf multi-index notation} as polynomials $\sum_k a_k X^k$ with non-negative 
rational integers $a_k=a_{k_1,\dots,k_n}$ and where $X^k=X_1^{k_1} * \cdots * X_{n}^{k_n}$ and where $X_k$ are both 
additive and multiplicative primes. Every integer $G \in \mathcal{Z}$ is of the form $G=A-B$ with 
$A,B \in \mathcal{N}$. This is unique if every connected component of $A$ is positive and every connected component of 
$B$ is positive. Unlike for the rational integers $\mathbb{Z}$, we can not always achieve that 
$G=A$ or $G=-B$. The graph $K_2-K_3$ for example
can not be simplified any further as there is no common additive factor. The multiplication extends from graphs
$\mathcal{N}$ to the {\bf group completion} $\mathcal{Z}$, so that $\mathcal{Z}$ becomes a 
{\bf commutative ring with $1$}. The {\bf empty graph} $0$ is the $0$-element and the one-point
graph $1 = K_1=\{ 0 \}$ is the {\bf one element}. The {\bf additive primes} are the connected components.
For the {\bf multiplicative primes}, we can skip the ``multiplicative" and simply call $\mathcal{P}$ the set of {\bf primes}.
$\mathcal{P} = \{ G \in \mathcal{Z} \setminus \{0,1\}, G=A*B \Rightarrow A=1$ or $B=1. \}$. 
The connected elements in this set are both multiplicative and additive primes. These are important building blocks because
every connected graph has a unique prime factorization into connected multiplicative primes.

\paragraph{\large{Extending quantities to integers and prime criteria.}}
The {\bf valuations} $f_k(G)$ counting the number of $k$-dimensional complete subgraphs in a graph $G \in \mathcal{N}$
can be extended to $\mathcal{Z}$ by setting $f_k(A-B)=f_k(A)-f_k(B)$. One then still has
$f_k(A+B) = f_k(A) + f_k(B)$ for all $A,B \in \mathcal{Z}$. One can so extend the $f$-vector from $\mathcal{N}$ to $\mathcal{Z}$.
Also $f_0(A*B) = f_0(A) \cdot f_0(B)$ and more generally $f_n(A*B) = \sum_{k+l=n} f_k(A) f_l(B)$ holds. 
The valuation $f_0(G)$ counting vertices is special as it is not only additive but also multiplicative. 
This allows us to see that if $G \in \mathcal{Z}$ has $f_0(G)$ as a 
rational prime (meaning to be prime in $\mathbb{Z}$), then $G$ must be prime in $\mathcal{Z}$. But there are
many more primes. An other integer-valued {\bf multiplicative number} is the {\bf clique number} $c(A)$ which gives the number
of vertices of the largest embedded $K_n$ in $A$. If the clique number of a graph $G$ is a rational prime, 
then $G$ is prime. Note that for graphs with clique number $1$, we do not have any edges and so deal with 
the standard arithmetic $\mathbb{Z}$. 
The multiplicative primes with clique number $1$ are the point graphs $P_p$ with $p$ vertices, where $p$ is prime.
As in standard arithmetic, also $-p$ is considered a prime if $p$ is prime. 

\paragraph{\large{Aspects of polynomial rings.}}
As custom in commutative algebra, it can be better to look at {\bf prime ideals} rather than primes. 
{\bf Principal prime ideals} are ideals that are generated by primes.
There are also plenty of {\bf maximal ideals}. These are automatically prime as in any commutative ring.
Since the ring $\mathcal{Z}$ contains any polynomial ring 
$\mathbb{Q}[x_1, \dots, x_n]$ as a sub-ring, one can get primes ideals from irreducible varieties.
Look at an irreducible polynomial $g$ and take the ideal $I=(g)$. It corresponds to an ideal in $\mathcal{Z}$. 
These are not maximal ideals but they define prime ideals in $\mathcal{R}$. In some sense, 
the arithmetic $\mathbb{R}$ contains structures usually considered in {\bf algebraic geometry}.
A function $p$ in $\mathbb{Q}[x_1, \dots, x_n]$ can be seen geometrically when choosing of prime graphs 
$X=(X_1,\dots,X_n)$ instead of variables. This then becomes a {\bf concrete network} $p(X) = p(X_1, \dots, X_n)$
once we get used to expressions like $-G$ or $1/G$ for prime graphs. 
We have a {\bf representation} of integer polynomial rings in a ring $\mathcal{Z}$
or $\mathcal{Q}$ of geometric objects. While this looks like a crazy overhead, there are interesting 
tasks like finding the {\bf prime factorization} of a graph $G$ or studying the geometric
properties complicated networks built as polynomials of smaller networks.

\paragraph{\large{The integers form an integral domain.} }
Let us again prove that the commutative unital ring of integers $\mathbb{Z}$ is an {\bf integral domain}.
For graphs $X,Y \in \mathcal{N}$ with vertex sets $V(X),V(Y)$,
the relation $X*Y=0$ implies $|V(X)| |V(Y)|=0$ so that $|V(X)|=0$ or $|V(Y)|=0$ and therefore $X=0$ or $Y=0$. 
Now let us look at at elements $X=A-B$ and $Y=C-D$ in $\mathcal{Z}$. 
If $(A-B)*(C-D)=0$, then $A*C+B*D=A*D+B*C$. Let us first assume that all graphs $A,B,C,D$ are all connected. 
Looking at the connected components we see $A*C=A*D$ or $A*C=B*C$. In the first case $A*C=A*D$ we have 
$C=D$ because the two graphs $A*C$ and $A$ determine $C$ uniquely by projection.
In the second case, one can in the same way conclude $A=B$. 
We have now shown the claim if $X=A-B,Y=C-D$ with connected $A,B,C,D$. We use this as an induction
base $n=1$ for the claim with $X,Y$ both having a total of $n$ connected components. 
Assume $X=A-B,Y$ where $A$ is connected and $B$ has one connected component less. 
Now $X*Y=(A-B)*Y=0$ means $A*Y=B*Y$. Again, the knowledge of $A*Y$ and $Y$ determines $A$ and
the knowledge of $BY$ determines $B$. 

\paragraph{\large{The integers do not form a unique factorization domain.} }
Also this has been mentioned in earlier write-ups or \cite{ImrichKlavzar,HammackImrichKlavzar}.
In order to see that we have not a unique factorization domain, we need an example. 
Take any connected positive-dimensional prime graph $X$ and form $A=(1+X+X^2)$, 
$B=(1+X^3)$ and $C=(1+X^2+X^4)$ and $D=(1+X)$. By looking at the vertex cardinality and noting that
for any integer $x$ larger than $1$ the $4$ numbers $1+x+x^2,1+x^3,1+x^2+x^4,1+x$ are different, 
we see that $A,B,C,D$ are different. Now assume $A=1+X+X^2= P Q$, then $P=1+U,Q=1+V$ with positive
dimensional $U,V$. The relation $X+X^2=U+V+U*V$ is not possible as there are different number of connected
components. Similar arguments hold for $1+x^3,1_x^2=x^4,1+x$ so that the $4$ graphs
$1+X+X^2,1+X^3,1+X^2+X^4,1+X$ are all {\bf multiplicative primes}. The phenomenon of non-uniqueness
only appears for disconnected graphs. A connected graph has a unique prime factorization. 
This is one of the first results proven in \cite{Sabidussi}.

\paragraph{\large{Number theory and factorization tasks.}}
We see that the arithmetic of integers $\mathcal{Z}$ produces number theory which goes far beyond 
the familiar number theory of rational integers $\mathbb{Z}$. The later is the {\bf $0$-dimensional part of 
$\mathcal{Z}$} which produces an {\bf arithmetic of signed graphs}. The {\bf additive primes} are the connected graphs, the 
{\bf multiplicative primes} are the graphs which can not be written as a product of two elements
different from $1$. Every additive prime $G \in \mathcal{N}$ can be written uniquely as a product 
$\prod_{j=1}^k P_j$ of multiplicative primes $\mathcal{P}$. If $P,Q$ are primes with $f_0(P)=n,f_0(Q)=m$
then $G=PQ$ is a graph with $f_0(G)=nm$. There are $2^{nm(nm-1)/2}$ graphs with $nm$ vertices and
$2^{n(n-1)/2}$ graphs with $n$ vertices so that only $2^{n(n-1)/2 + m(m-1)/2}$ composite numbers among
all $2^{nm(nm-1)/2}$ graphs. This shows that almost all graphs are prime as the set of products is thin.
A really interesting task for a computer scientist is to see {\bf how effectively one can factor a network }
$A*B$. By looking at multiplicative functionals like Euler characteristic, Wu characteristic,
clique number, total number of cliques, number of connected components
or number of vertices, we can get clues about how these quantities   
look like for $A,B$. 

\paragraph{\large{The choice of localization.}}
The concept of {\bf localization} of a ring $R \to S^{-1} R$ with respect to a multiplicatively closed set $S$ 
in a ring allows to extend $R$ so that we can divide by elements in the monoid $S$. A standard example is to 
get from a polynomial ring $K[x]$ to a {\bf Laurent polynomial ring} $K[x,x^{-1}]$. Localization preserves
integral domains. An other example which works in integral domains
is to divide out the entire multiplicative monoid leading to the {\bf field of fractions}.
The name {\bf localization} comes from the fact that for a field $K$, when taking $K[x]/(x-a)$ we get the field $K$ back
and localize it in some sense at the point $a$. When done with the ring $\mathcal{Z}$ and taking the 
multiplicative monoid of expressions generated by multiplicative and additive primes $G$, 
we get the ring $\mathcal{Q}$ of {\bf rational numbers}. The choice of localization 
is justified when doing the completion which needs to preserve a norm and for which one can not choose
a too large monoid. The classification of division algebras shows that we can not get a field for example
which carries a Banach norm as this would give a new Banach space which is also a field and so has to be either
the traditional real of complex numbers $\mathbb{R}$ or $\mathbb{C}$.

\paragraph{\large{Laurent polynomials}}
The full ring $\mathcal{Q}$ is isomorphic to the ring of {\bf Laurent polynomials}
with any number of variables in $\mathcal{P}$.
This is justified by the fact that every connected component has a unique prime factorization.
Once we have that, we can look at a norm and complete the ring to get the ring $\mathcal{R}$ of real numbers.
Let $\mathcal{S}$ denote the multiplicative monoid generated by $\mathcal{P}$ the both 
multiplicative as well as additive primes. This is a countable set and quite large: for example, every 
connected graph $G$ with prime cardinality $|V(G)| \in \mathbb{P}$ 
must be in $\mathcal{P}$ because 
$V(G_1*G_2)=V(G_1) \times V(G_2)$ forces connected graphs with prime cardinality to be prime. 
The localization $\mathcal{Q} =\mathcal{S}^{-1} \mathcal{Z}$ can now be seen as the set of
all Laurent polynomials in the countable set of variables $\mathcal{P}$. 

\paragraph{\large{Rational numbers still form an integral domain.}}
The {\bf classical rational numbers} $\mathbb{Q}$ are strictly contained in the {\bf field of 
fractions} $\mathcal{F}=\mathbb{S}^{-1} \mathbb{Z}$, 
which is the largest possible localization as it uses the multiplicative monoid $\mathcal{S}$ of non-zero 
elements in $\mathbb{Z}$ and so allows to define $A/B$ for every non-zero $B$. This is a field $\mathbb{Q}$,
the field of rational numbers. 
The rational numbers $\mathcal{Q}$ still form a commutative unital ring over the field of 
rational numbers $\mathbb{Q}$. It is still an integral domain
because localization in general preserves integral domains.
The property of having an integral domain will disappear in the Banach algebra $\mathcal{R}$. 
The reason is that in the one-graph network $\mathcal{R}[G]$ where we get continuous functions on the circle 
with absolutely convergent Fourier series, we can have an element $A$ which is zero on half of the circle 
and an other element $B$ that is zero on the other half of the circle. The product $A*B$ is zero 
but none of the networks $A,B$ are zero. This insight is one o the reasons, why the Wiener picture is important. 

\paragraph{\large{Looking at irrational and non-algebraic numbers.}}
We can identify $\mathcal{Q}$ as a set of Laurent polynomials using a countable set of
variables $\mathcal{P}$, where each variable is in the set of additive and multiplicative primes $\mathcal{P}$. 
The number $X=1/(5-K_2)=(1/5)/(1-K_2/5)$ is then only defined in the completion $\mathcal{R}$ and not in $\mathcal{Q}$. 
It is not a rational number but already an {\bf irrational number}. We need an infinite sum
$5^{-1} \sum_{k=0}^{\infty} K_2^k/5^k$ to represent it but is a well defined positive real number in $\mathcal{R}$. 
We could have extended the localization process to include a larger set $\mathcal{S} \subset \mathcal{Z}$ but 
that requires that we know which elements in $\mathcal{Z}$ are ``positive". Going with Laurent polynomials will get us 
for one network in the limit to the familiar Wiener algebra of continuous functions on $\mathbb{T}$ for which we know
what the invertible elements are: they are the double cone $\{ f >0 \} \cup \{ f < 0 \}$ in the Wiener
algebra $A(\mathbb{T})$. One can also mimic Cantor's proof for the existence of non-algebraic numbers by just 
looking at the cardinality. If an {\bf algebraic number} is a solution to a polynomial equation $p(X)=0$ where the
coefficients of $p$ are in $\mathcal{Z}$, then there are only a countable set of algebraic numbers but an uncountable
set of numbers in $\mathcal{R}$. 

\paragraph{\large{Subrings generated by finitely many networks.}}
Instead of looking at the full ring $\mathcal{Q}$, one can look at subrings in which we only adopt a class
of graphs. Such a restriction can make sense in a computer science setting, where we are maybe only interested
in a few networks, or even in only one network. The simplest case is if we work with a single network $X$ only. 
The ``rational numbers" then are as a set isomorphic to $\mathbb{Q}[X,X^{-1}]$ within $\mathcal{Q}$. 
An ultra-finitist can still be happy with an arithmetic in which a {\bf finite set of networks} is 
allowed for the computation. In that case, we have a Wiener algebra with a finite dimensional 
set of maximal ideals. We could fix notation by 
calling $\mathcal{Q}[G]$ the ring $\mathbb{Q}[G,G^{-1}]$ of Laurent polynomials in one variable
and $\mathcal{Q}[G_1, \dots, G_n]$ the ring generated by Laurent polynomials of several variables.
This ring is {\bf generated} by the product of {\bf one particle Laurent polynomial rings} but not
equal to the product. Most expressions in the ring can not be written as products of expressions in 
the individual networks. These are {\bf entangled rational numbers}. Lets look at $\mathcal{R}[A,B]$
which contains expressions of the form $G=\sum_{n,m} a_{n,m} A^n B^m$. Most of these expressions are
not products of expressions in $\mathcal{R}[A]$ and $\mathcal{R}[B]$. 

\paragraph{\large{Multiplicative functionals}}
A map $\phi: \mathcal{Q} \to \mathbb{C}$ is a {\bf multiplicative functional} 
if $\phi(1)=1,\phi(AB)=\phi(A) \phi(B)$. It is an {\bf additive functional} (also called {\bf linear functional}
or {\bf valuation}) if $\phi(A)+\phi(B)=\phi(A)+\phi(B)$. If a functional is both multiplicative and additive,
then it defines {\bf ring homomorphism} $\mathcal{Q} \to \mathbb{C}$. One can then look 
at the kernel $\{ G, \phi(G)=0 \}$. It is a geometric space which is invariant under addition and multiplication.
In a geometric setting, there are many natural examples already. For the Euler characteristic $\chi(G)$ one
gets as the kernel the set of graphs with $0$ Euler characteristic. This space is invariant under both addition
and multiplication. An other functional is $f_0(G)$ counting the number of vertices in $G$.
It is a ring homomorphism from $\mathcal{Z}$ to $\mathbb{Z}$. The kernel consists of all graphs in $\mathcal{Z}$ for which 
the number of vertices of positive graphs is the same than the number of vertices of negative graphs.
In general, note that if a multiplicative functional $\phi: \mathcal{Z} \to \mathbb{Z}$ has the property that $\phi(G)$ is 
a rational prime and $\phi(G)=1$ implies $G=1$, then $G$ is a multiplicative prime. 

\paragraph{ \large{Euler characteristic and Wu characteristic.}}
The {\bf Euler characteristic} $\chi(G)$ is an other both multiplicative and additive functional
from $\mathcal{Z}$ to $\mathbb{Z}$. It first must be extended from $\mathcal{N}$ to $\mathcal{Z}$ by the assumption
$\chi(-A)=-\chi(A)$. The Euler characteristic extends then to $\mathcal{Q}$ by defining $\chi(A^{-1})=1/\chi(A)$ for 
any connected component $A$. It can not be extended to all of $\mathcal{R}$. 
While sometimes, it produces a finite answer like $G=\sum_k B^k/c(B)^k$ 
$\chi(B)<c(B)$, this fails in general $|\chi(B)|>|B|$, then the Euler characteristic $\chi(G)$ of the limiting
object $G$ is not defined. The {\bf Wu characteristic} is 
$\omega(A) = \sum_{x \sim y} \omega(x) \omega(y)$
with $\omega(x) = (-1)^{{\rm dim}(x)}$ summing over all pairs of complete subgraphs $x,y$ of $A$.
Which functions we can extend to $\mathbb{R}$ depends on the radius of convergence. There is no problem 
to define the Euler characteristic for elements like $\exp(X) \in \mathcal{R}$. 
We have $\chi(\exp(X)) = \exp(\chi(X))$ and $\omega(\exp(X)) = \exp(\omega(X))$ for example.

\paragraph{\large{Extending spectral zeta function to real networks}}
The {\bf spectral zeta function} $\zeta_G(s)$ defined by eigenvalues of the connection Laplacian $L_G^2$ 
(which is a positive definite matrix) is a both
an additive and a multiplicative functional.  To the definition:
given a connection matrix $L_G$ of $G$, we can look at the 
{\bf zeta function} $\zeta_G(s) = \sum_{k} \lambda_k^{-s}$, where $\lambda_k$ are the 
eigenvalues of $L_G^2$. The reason for taking the square is to get positive eigenvalues to assures
that $\lambda_k^{-s}$ is defined for all complex $s$. For any fixed $s$, the functional $A \to \zeta_A(s)$ 
defines a both additive and multiplicative functional $G \to \zeta_G(s)$.
What is nice in $\mathcal{R}$ is that we can define $\zeta_G(s)$ as long as $G = f(A)$ is a 
``real number" which is obtained by an entire function $f$. We can for example look at 
the zeta function $\zeta_{e^{i A t}}(s)$ of the wave $e^{i At} = \cos(At) + i \sin(At)$. 
It is simply given by $e^{i \zeta_A(s) t}$. We should add that while in $C^*$-algebras, the functional 
calculus can be pushed to {\bf all continuous functions}, but in a Banach algebra like ours $\mathcal{R}$, 
we can a priory only use an analytic functional calculus. 

\paragraph{\large{The picture of maximal ideals}}
As in any unital Banach algebra $\mathcal{R}$, a maximal ideal $m$ defines a {\bf field} $\mathcal{R}/m$. 
The algebra $\mathcal{R}$ is huge and contains many maximal ideals. Since it contains products of Wiener algebras
the space of maximal ideals naturally contains the topological space $\mathbb{T}^{\infty}$.
These correspond to the {\bf multiplicative linear functionals} in the product of all subalgebras $\mathcal{R}[G_i]$ 
which is a strict subset of $\mathcal{R}$ but generate $\mathcal{R}$. 
In many cases, we might be interested in one network only. In that case the set of maximal ideals is the circle.
For any $t \in \mathbb{T}$ one has the functional $\phi_t(G) = w_G(t)$, where
$w_G(t)$ is the continuous function on $A(\mathbb{T}$ which corresponds to the element $G$. 
How can one get the function $f$ which belongs to an element
$X=\sum_{n=-\infty}^{\infty} a_n G^n \in \mathcal{R}$? The answer is simply 
$f(t)=\sum_{n=-\infty}^{\infty} a_n e^{i n t}$. As $|X|=\sum_{n} |a_n|$ is finite, we the function
$f(t)$ is well defined. For every differential periodic function and a choice of base network $G$,
one has an associated element $X=\sum_{n=-\infty}^{\infty} a_n G^n \in \mathcal{R}$. But of course, 
in order to do the multiplication of any two elements $X,Y$, one just multiplies the corresponding 
periodic functions in $A(\mathbb{T})$. The multiplicative linear functionals on the sub algebra are
then just the evaluations of the function at a point $t$. 

\paragraph{\large{Extending connection graphs to the real numbers.}}  
For $G \in \mathcal{N}$, the {\bf connection graph} $G'$ is defined by the edge set $E=\{ (a,b), a \cap b \neq \emptyset \}$.
For the Cartesian product $G=G_1 \times G_2$, the {\bf connection graph} is
defined as  $(G_1 \times G_2, \{ ( (a,b), (c,d) ), a \cap c \neq \emptyset, \; {\rm and} \; b \cap d \neq \emptyset \})$. 
This is the {\bf strong product} of the connection graphs $G_1'$ and $G_2'$.
The Cartesian product produces the strong product on the level of the connection graph and defines the tensor product
for the connection Laplacians $L_A$ which is in the simplest case defined as $L_A(x,y)=1$ if $x,y$ intersect and 
$L_A(x,y)=0$ if not. For $G=A-B$, just define $G'=A'-B'$ in $\mathcal{Z}$. We can extend this to $\mathcal{Q}$ by 
defining for $A,B,C \in \mathcal{N}$ the expression
$((A-B)/C)' =(A'-B')/C'$. Connection graphs are so defined for every $G \in \mathcal{Q}$.
We have seen that the linearity and multiplicative property of the zeta function to extend 
also the zeta function to $\mathcal{Q}$ and even to part of $\mathcal{R}$. The point we wanted to make here
that it can depend on the norm whether $f(G')$ still makes sense. For entire functions $f$, 
there is never a problem to define $f(G)'$. 

\paragraph{\large{Proving the Banach algebra property}}
The proof of the normed ring property can help to see which choice of multiplicative functional do work.
Given such a function $c(G)$, then define the norm $|G| = \inf_{G=A-B} c(A)-c(B)$
satisfies the {\bf Banach algebra property} $|A*B| \leq |A| |B|$ for $A,B \in \mathcal{Q}$. 
Proof: For $A,B \in \mathcal{N}$ we have $c(A * B) = c(A) c(B)$ and so $|A*B|=|A| |B|$. However,
$|(A-B)(C-D)| = c(AC+BD) + c(AD+BC)$ is smaller than $(c(A)+c(B))(c(C)+c(D))=
c(AC) + c(BD) + c(AD) + c(BC)$ if one of the $A-B$ and $C-D$ is not either positive or negative
or zero. This works for the clique number because 
$c(X+Y)={\rm max}(c(X),c(Y)) \leq c(X) + c(Y)$ and that we have strict inequality 
if not one of the two $X,Y$ is $0$. 

\paragraph{\large{The Wiener space}}
The Banach algebra $\mathcal{R}$ contains the product of Wiener algebras $\mathcal{R}[P]$ for every 
additive and multiplicative primes. Therefore, $\mathcal{R}$ contains the larger Wiener space $A(\mathbb{T}^{\infty})$ 
of all continuous functions on the infinite dimensional torus which have an absolutely convergent
Fourier expansion. But this is not yet the entire $\mathcal{R}$. While the product Wiener space {\bf generates}
$\mathcal{R}$, it is not equal to $\mathcal{R}$.  Already the space $\mathcal{R}[A,B]$ generated by two networks
$A,B$, contains the product space $\mathcal{R}[A] \times \mathcal{R}[B]$. The later generates $\mathcal{R}[A,B]$.
One could also ask about the completion and analogue of $\mathcal{C}$ even in a one-network Banach algebra.
This is hopeless for Banach algebras. Already the completion of $\mathcal{B}[G]$, the Wiener algebra is
large as already $f X = g$ does in general not have a solution $X$ if $f$ has roots. 

\paragraph{\large{Different completions.}}
Let us look at the one particle case $\mathbb{R}[G]$ in which one network $G$ extends the real numbers $\mathbb{R}$.
It is isomorphic to the
{\bf Wiener algebra} $A(\mathbb{T}^{\infty}) \subset C(\mathbb{T}^{\infty})$. 
When completing the rational numbers it depends, what norm we chose for the primes. For example, when 
choosing $c(P)=1$ for every $P \in \mathcal{P}$ and extending this to products results in $|A-B| =2$
if $A,B$ are monoids. A concrete way to represent elements in $\mathcal{Q}$ is to write them like 
elements in $\mathbb{R}[P_1,P_2,\dots ]$ where $P_i \in \mathcal{P}$. 
These are sums of terms $a_k P^k = P_{1}^{k_1} \cdots P_n^{k_n}$ using multi-index notation.
The norm is $|G| = \sum_{k \in \mathbb{Z}^{\mathcal{P}}} |a_k| c(P)$ whatever positive number choice $c(P)$ 
has been made on primes.  Elements $\hat{G}$ in the Wiener algebra 
$A(\mathbb{T}^{\mathbb{Z}}) \subset C(\mathbb{T}^{\mathbb{Z}})$ can be written then as
$\hat{G} = \sum_{k \in \mathbb{Z}^{\mathcal{P}}} a_k e^{i k \cdot x}$, 
where $x=x_{p \in P} \in \mathbb{T}^{\mathcal{P}}$ represents a point on the infinite dimensional torus. 
The multiplication in Gelfand-Fourier space is the point-wise computation.

\paragraph{\large{Questions about primes.} }
It is important to ask how difficult it is to factor a given $G \in \mathcal{N}$. (This could have
practical applications when trying to build crypto systems based in $\mathcal{Z}$).
If a graph $G$ has $n$ vertices, we need to find two subgraphs $A,B$ of size $p,q$
such that $pq=n$ and such that $A*B=G$.
This can also be asked in sub-algebras. Any class of graphs
defines an algebra and we can ask whether we can decompose a graph within that class.
For example, in the class of complete graphs, the primes are $K_p$, where $p$ is a rational prime
because $K_p*K_q=K_{p*q}$. 
We could also look at all graphs in $\mathbb{Z}$ which are connection graphs. Is it true that 
if $G$ is prime in this subalgebra, then $G$ is prime? 
There are many primes like complete graphs $K_{p}$ with prime $p$ which are prime graphs 
in $\mathcal{Z}$ but not prime connection graphs.
So there could in principle be possible to factor a connection 
graph using non-connection graphs. 

\paragraph{\large{About algebraic completion.}}
Let us work in  the ring $\mathcal{R}[G]$ of one network which is isomorphic to the
Wiener algebra $A(\mathbb{T},\mathbb{R})$. We can certainly also extend this to the 
complex numbers $\mathbb{C} =\mathbb{R}[i]$ and get the 
{\bf complex Wiener algebra} $\mathcal{C}[G] = A(\mathbb{T},\mathbb{C})$. 
We can not find additional solutions of the equation $X^2+1=0$. The 
reason is that this equation does not have solutions in the Banach algebra unless $X$ is constant
because the elements in the Wiener algebra are continuous functions. [Note that we can not argue just
algebraically because $\mathcal{R}[G]$ is far from an integral domain. There is a big difference
between analytic functions and smooth functions: for analytic functions $fg=0$ implies $f=0$ or $g=0$
but for smooth functions this is not the case. There are even $C^{\infty}(\mathbb{T})$ functions $f,g$
which are non-zero but for which $fg=0$. ] More generally, any solution $p(G)=0$ with a polynomial
$p$ having complex $\mathbb{C}$ coefficients has only the solutions in $\mathbb{C}$ and not in the larger
algebra $\mathcal{C}[G]$. The argument is the same. As the algebra is given by continuous functions
and at every point only the solutions of $f(z)=0$ can occur as values, the function has to be constant. 
This extends to the full algebra $\mathcal{R}$ (or $\mathcal{C}$) even so these algebras are not 
integral domains. As for polynomials with coefficients in $\mathcal{R}$ this is hopeless: 
we can not solve already linear equations $f X = g$ as we have seen that $f$ is invertible only 
if it has no roots. A completion of $\mathcal{R}$ would be huge.

\paragraph{\large{Exponentiation of integer networks.}}
Given two networks $B,X$. How do we define $B^X$ most naturally? We can try to 
write $e^{X \log(B)}$ and use power series for $B=1+b$. This works only if the norm is chosen
so that $b=1-B$ can have norm smaller than $1$. With the given norm so that integers $A \in \mathcal{N}$
have integer norm, this does not work. The approach works if $B>0 \in \mathbb{R}$ because in that
case we know what the logarithm is for all $B$. To define $B^X$ in that case, just build a Taylor expansion of $f(x)=B^x$ 
and plug in $X$. As $f$ is entire function, the functional calculus goes through in the Banach algebra already.
With an analytic approach failing in general, we can ask for operations having 
the usual properties $(B C)^X = B^X C^X$, $1^X=1$, $B^{X+Y}=B^X B^Y$. 
The simplest way to achieve this is to take a linear multiplicative functional $\phi$ on graphs to the natural 
numbers and then define $B^X = B^{\phi(X)}$. We originally were using connection calculus
to do that and define $B^X = {\rm det}(L_h(X))$ where the energization $h$ attaches to each simplex $x$ the value 
$B^{|x|}$. But then since ${\rm det}(L_h(X))$ is just $B^{\phi(X)}$ where $\phi(X)$ counts the total number
of simplices in $X$. An other choice could be $B^X=B^{\chi(X)}$, where $\chi$ is the Euler characteristic. 
So, while we can define $B^X$ for $B,X \in \mathcal{N}$ in different ways, we do not know what is the most ``natural" one.
The ambiguity is not unexpected since already the definition of $B^X$ for complex $B,X$ has ambiguous due to the fact
that there are infinitely many logarithms which work. 

\paragraph{\large{Is there an Ostrowski theorem or a p-adic Wiener theorem?}}
One can ask for an analog of {\bf Ostrowski's theorem} to get all possible valuations on $\mathcal{Q}$ and
so topological completions of $\mathcal{Q}$ to something larger. 
We seem to have a lot of choice already to choose a Banach
norm on $\mathcal{Q}$. Most likely there are also lot of different norms which are not equivalent. 
There is a p-adic norm possibility by taking a rational prime $p$ and then look at the $p$-adic norm 
$|x|_p$ on the usual rationals $\mathbb{Q}$ which when completed gives the $p$-adic numbers $\mathbb{Q}_p$. 
Let us attach a single network $G$ and look first at $\mathcal{Q}[G] = \{ X = \sum_{k=-m}^n a_k G^k \}$
with norm $|X| = \sum_k |a_k|_p$, then look at the completion which is the $p$-adic Wiener algebra 
$l^1(\mathbb{Z},\mathbb{Q}_p)$ of $l^1$ sequences taking values in the field of p-adic numbers rather than 
in the field of real or complex numbers as usual. Here is a question for p-adic harmonic analysis:
while before the Wiener algebra was $A(\mathbb{T})=A(\mathbb{R}/\mathbb{Z})$
and $\hat{\mathbb{Q}_p} \sim \mathbb{Q}_p$ (Tate's theorem) and $\hat{\mathbb{Z}_p}=Z(p^{\infty})$ 
is the Pr\"ufer group that $l^1(\mathbb{Z},\mathbb{Q}_p)$ is isomorphic to 
$A(\mathbb{Z}_p) = A(\mathbb{Q}_p/\mathbb{Z})$ the set of continuous functions on the p-adic integers
for which the sum of the absolute values of all Fourier coefficients in $Z(p^{\infty}$ is finite. 

\paragraph{\large{About Gelfand's proof of the Wiener theorem.}} 
The proof of Wiener's $1/f$-theorem works also higher dimensions where one has a Wiener algebra
$A(\mathbb{T}^n)$ consisting of continuous functions on the $n$-torus $\mathbb{T}^n$. 
The proof of Gelfand is the most
elegant. First check that the product of absolutely convergent Fourier convergent
functions on $\mathbb{T}^n$ produce the convolution $c_n = \sum_{k+l=n} a_k b_l$, then 
check the Banach algebra property $\sum_n |c_n| \leq \sum_{k+l=n} |a_k| |b_l|$
which is $\sum_k |a_k| \sum_l |b_l|$ showing the Banach algebra property. If an 
element $f$ is not invertible, then $\{ f x, x \in A(\mathbb{T}^n) \}$ is not the 
entire ring and contained in a maximal ideal which is of the form $\{ f, f(\tau)=0 \}$
for some $\tau \in \mathbb{T}^n$. So, if $f$ is not invertible
then $f$ has a zero. If $f$ has nowhere zero, it is invertible. 

\paragraph{\large{A nonstandard consideration.} }
The following set-up uses a picture familiar in nonstandard analysis like internal set theory.
If $M$ is a compact topological space, there is a finite set $V$ which has the property that every $x \in M$
is infinitesimally close to an element in $V$. The notion ``infinitesimal" is defined
through axioms like being smaller than any positive standard number. Now given
any infinitesimal $\epsilon>0$ we have a graph $(V,E)$, where
$E=\{ (x,y), d(x,y)<\epsilon \}$. Given such an $\epsilon$ graph $G_k=(V_k,E_k)$ of
compact spaces $M_k$, the strong product $G_1*G_2$ is an $\epsilon$ graph for
$M_1 \times M_2$ if we take the taxi product metric. We see that the Shannon product 
appears quite naturally if we approximate manifolds by graphs encoding the topology. 

\paragraph{\large{Phenomenology of integers.}}
Euclid thought about numbers as a ``multitude of units" (``To ek monadon synkeimenon plethos").
As discussed by phenomologists like \cite{HusserlArithmetic,Miller1982}, numbers are usually not seen 
geometrically. Husserl: {\it ``So it is clear that number concepts and relationships between numbers - much more so the whole of 
arithmetic - have nothing to do with the representation of space."} 
Our relations with number still are very much based in physics. 
Many kids have a physical relation to numbers, for example through Cuisenaire material (Georges Cuisenaire 1891-1976).
There are strong spacial and temporal connections with numbers like with Khipu
\cite{UrtonInkaHistoryKnots}. Nature has encoded information in DNA. Our thinking is fundamentally linked to 
molecule arrangements and computations as long term memory in our brain is encoded in brain cell like DNA methylation.
It is no surprise that various aspects and philosophical ideas about numbers exist.
An computer science point of view is Zeilberger's {\bf ultrafinite computerism} \cite{Zeilberger2001}:
{\it "So I deny even the existence of the Peano axiom that every integer has a successor.
Eventually we would get an overflow error in the big computer in the sky. }
An ultra finitist would therefore rather work in the Laurent ring $\mathcal{Q}$ and not use
any completion $\mathcal{R}$. Still, as explained by Zeilberger, being a finitist does not prevent us to 
look at expressions like $G=e^{i A t}$ for a network $X$, even so we in principle need infinitely many terms
in a Taylor expansion to define it. We can just look at $G$ as an object like a complex number $e^{i a t}$ has
been looked at before as an expression we can compute as accurately as we need it given the limitations we have.

\paragraph{\large{Number theoretical questions.}}
Every number theoretical question asked in $\mathbb{N}$ can be asked in $\mathcal{N}$. Some of them are interesting
in the more geometric setup of networks, some are easy to answer and others directly relate to the questions in $\mathbb{N}$.
The infinitude of primes for example follows directly from $\mathbb{N} \subset \mathcal{N}$ or that if $\phi$ is a multiplicative
functional then a prime $\phi(G)$ assures $G$ is prime. The {\bf partition problem} to compute $p(G)$, the number of times
that a graph $G$ can be written as a sum of smaller non-empty graphs is just $p(b_0(G))$, where $b_0(G)$ is the number of 
components. There is nothing more interesting in that respect. 
Other questions just do not work because there are too many elements. 
An $\sum_{G} G^{-s} = \prod_{P} (1-P^{-s})$ for graphs does not even work for any $s$ if we bound the clique number.
(The left sum for example is over all connected graphs of dimension $2$ or less and the right hand side sums over
all connected primes in this class). 

\paragraph{\large{Additive questions about primes.}}
While {\bf Landau's problem} whether there are
infinitely many primes of the form $n^2+1$ is open, one can say that for all connected networks $A$,
the graph $G=A^2+1$ is prime simply because the number of connected components is $2$ which is prime. More generally for
any $n,m$ and any connected graphs $A,B$, the graph $G=A^n+B^m$ prime. 
The analogue of the Goldbach conjecture is not interesting because if $G$ is an additive prime (connected) 
that is not a multiplicative prime, then $N=2G$ can not be the sum of two primes because the only way to 
write $N=2G$ is $N=G+G$ but then $G$ was assumed to be not prime.
There are no multiplicative primes of the form $P+Q$, where $P,Q$ are two additive primes because the connected
components of $P+Q$ is $2$, which is prime.

\paragraph{\large{How many graphs with n nodes are prime?}}
How does the number of multiplicative primes of graphs with $n$ vertices grow in $\mathcal{N}$? 
Among all the $2^{n(n-1)/2}$ possible graphs with $n$ vertices, {\bf most are prime}.
If $n$ is prime, then even {\bf all of them are prime}. 
In general, the set of multiplicative primes make up almost all integers $\mathcal{N}$
simply because if $n=pq$, then the number of graphs with $n$ vertices is so much larger than the product of the number of graphs
with $p$ and $q$ vertices gives at most $2^{p(p-1)/2} \cdot 2^{q(q-1)/2}$ factors (not even requiring them to be prime).
Since $p*q=n$ and $p,q \geq 2$, we have at most $2^{n/2 (n/2-1)/2} 2$ non-primes. 
The probability to get a prime graph among all graphs with $n$ nodes is therefore smaller than $2^{1+n/4-3n^2/8}$. 
One could ask for the probability in Erd\"os-Renyi probability spaces $E(n,p)$ of graphs. In $E(n,0)$ most graphs are 
non-prime, in $E(n,1)$  most graphs are prime. 

\paragraph{\large{Arithmetic progressions}}
One can also ask the analogue of Dirichlet's theorem on arithmetic progressions $A+n G$, where $A,G$ are graphs. 
Again, by looking at ring homomorphisms $\phi$ like the number of connected components $b_0(G)$ or the number of vertices 
$f_0(G)$ or Euler characteristic $\chi(G)$ one can get infinitely many primes in the sequence $A+nG$ as long as
$\phi(A)$ and $\phi(G)$ are co-prime non-zero numbers, then the statement 
for $\mathcal{N}$ follows from the Dirichlet theorem in $\mathbb{N}$. But we can even ask and more and require that all but finitely 
many elements in $A+nG$ to be prime. For example, all graphs of the form $1+n*G$ with connected positive dimensional
$G$ are either $1$ or prime. The reason is that any factor must be of the form $1+A$ with positive dimensional $A$
but that a product like $(1+A)(1+B) = 1+A+B+A*B$ produces connected components with more than 2 clique numbers.
It seems therefore that there is an {\bf anti-Dirichlet theorem} for graphs stating that in any arithmetic sequence $A+nG$
with $A,G$ co-prime connected positive dimensional graphs, there are only finitely many non-primes. 
We do not have examples yet of a pair of co-prime connected positive dimensional graphs where infinitely many $A+nG$ are
composite. 

\paragraph{\large{The factoring problem}}
Given a graph $G=AB$, where $A,B$ are connected primes. How do we reconstruct $A,B$. We have clues from multiplicative
functions like clique number or Euler characteristic. An other clue comes from a general Levitt relation for curvature
$$ K(v) = \sum_{k=0}^{\infty} \frac{(-1)^k V_{k-1}}{(k+1)} 
        = 1 - \frac{V_0}{2} + \frac{V_1}{3} - \frac{V_2}{4} + \cdots  \; , $$
where $V_k(v)$ is the number of $k+1$-dimensional simplices containing $v$ \cite{cherngaussbonnet}. 
This curvature for any graph $G=(V,E)$ satisfies $\sum_{v \in V} K(v) = \chi(G)$, the Euler characteristic.
What happens is that for a vertex $(v,w) \in V(A \star B)$ one has $K(v,w) = K(v) K(w)$. Since the 
$K(v)$ are rational numbers this could be used to find the factorization.
The product property can be proven with integral geometry. Here is a sketch: $K(v)={\rm E}[i_f(v)]$ is the expectation
of Poincar\'e-Hopf indices $i_f(v)=1-\chi(S^-_f(v))$, where $S^-_f(v)$ is the part of the unit sphere $S(v)$
where $f(w)<f(v)$. We have $i_{f,g}(v,w) =i_f(v) i_g(w)$ because of independence and because
the unit sphere (as well as stable unit spheres) of $S_{f,g}(v,w)$ become the homotopic to the join of $S_f(v)$ 
and $S_g(w)$.

\bibliographystyle{plain}

\end{document}